**Epitaxial growth of topological insulator $Bi_2Se_3$ film on Si(111) with atomically sharp interface**

By *Namrata Bansal[1], Yong Seung Kim[1], Eliav Edrey, Matthew Brahlek, Yoichi Horibe, Keiko Iida, Makoto Tanimura, Guo-Hong Li, Tian Feng, Hang-Dong Lee, Torgny Gustafsson, Eva Andrei*, and *Seongshik Oh[\*]*

[*]  S. Oh
Department of Physics and Astronomy
Rutgers, the State University of New Jersey
Piscataway, NJ 08854 (USA)
E-mail: ohsean@physics.rutgers.edu

N. Bansal
Department of Electrical and Computer Engineering
Rutgers, the State University of New Jersey
Piscataway, NJ 08854 (USA)

Y. S. Kim
Graphene Research Institute
Sejong University
Seoul 143-747 (Korea)

E. Edrey, M. Brahlek, Y. Horibe, T. Feng, G-H. Li, H-D. Lee, T. Gustafsson, E. Andrei
Department of Physics and Astronomy
Rutgers, the State University of New Jersey
Piscataway, NJ 08854 (USA)

K. Iida, M. Tanimura
Research Department, Nissan Arc, Ltd.
Yokosuka, Kanagawa 237-0061 (Japan)

[1]: Authors contributed equally to the paper.

(Last modified: April 17[th] 2011)

Atomically sharp epitaxial growth of $Bi_2Se_3$ films is achieved on Si (111) substrate with MBE (Molecular Beam Epitaxy).   Two-step growth process is found to be a key to achieve interfacial-layer-free epitaxial $Bi_2Se_3$ films on Si substrates.   With a single-step high temperature growth, second phase clusters are formed at an early stage.   On the other hand, with low temperature growth, the film tends to be disordered even in the absence of a second phase.  With a low temperature initial growth followed by a high temperature growth, second-phase-free atomically sharp interface is obtained between $Bi_2Se_3$ and Si substrate, as verified by RHEED (Reflection High Energy Electron Diffraction), TEM (Transmission Electron Microscopy) and XRD (X-Ray Diffraction). The lattice constant of $Bi_2Se_3$ is observed to relax to its bulk value during the first quintuple layer according to RHEED analysis, implying the absence of strain from the substrate. TEM shows a fully epitaxial structure of $Bi_2Se_3$ film down to the first quintuple layer without any second phase or an amorphous layer.

## 1. Introduction

Topological insulators (TI) are emerging as an ideal platform for spintronics and other applications such as quantum computations over the past few years[1-4]. Integrating TI with the mature Si technology is of key importance in order to realize its full potential for future devices. Amongst the materials being pursued as TI, $Bi_2Se_3$ stands out because it has almost the largest bulk band gap of 0.3eV, and a simple surface state structure of a well-defined single Dirac cone at the momentum zero point in k-space[5-9]. Though efforts have been made to grow single-crystalline $Bi_2Se_3$ on Si, the growth was accompanied by disordered $Bi_2Se_3$ layers at the interface [10-11], and an atomically sharp interface between $Bi_2Se_3$ and Si has not been reported yet; one of the main problems being the reactivity of the Si surface. Here we report an atomically sharp epitaxial growth of $Bi_2Se_3$ films on Si(111) substrate without a disordered interfacial layer; a proper substrate preparation step and two-temperature growth process were the key.

$Bi_2Q_3$ (Q= Se, Te) compounds and their solid solutions have been studied for a long time because of their outstanding thermoelectric properties [12-13]. However, following the theoretical prediction about $Bi_2Q_3$ being a topological insulator, interest has been renewed in this material. Topological insulators are characterized by an insulating bulk state with a band gap and metallic surface states protected by time reversal symmetry; the surface states being immune to nonmagnetic impurities or disorders. Strong spin-orbit coupling in heavy element, small gap semiconductor materials, causes carriers to feel a spin-dependent force even without external magnetic field, and forces the spin orientation of the surface carriers to lock to their flow direction, thus preventing backscattering from occurring on the metallic surface state [1-5,14-16]. This phenomenon of spin-momentum locking and the existence of a single Dirac cone were first observed on cleaved bulk samples using angle resolved photo emission spectroscopy (ARPES) [9]. Several other predicted TI properties such as Landau levels measured by STM in bulk [17] and thin films [5], existence of a metallic surface state as observed through Ahronov-Bohm effect[8], observation of weak anti-localization behavior [18], and Shubnikov de Haas oscillations [19], etc. have also been reported.

The sequence of Se-Bi-Se-Bi-Se forms a stack or quintuple layer (QL) along the c-axis direction of the $Bi_2Se_3$ rhombohedral structure. The bonding between different quintuple layers is weak van der Waals type, commonly associated with layered structures. However, the bonding within the stack or the Bi-Se bonding is stronger [12,20], enabling the film to grow QL-by-QL. Earlier methods to prepare $Bi_2Se_3$ films, such as electrodeposition [21-22], solvo-thermolization [23], thermal evaporation [24], successive ionic layer adsorption and reaction[25], chemical bath deposition [26] etc. generally produce polycrystalline films; though suitable for thermoelectric studies, these are unsuitable for study as a thin film topological insulator; thus, making a need for a single crystal $Bi_2Se_3$ thin film acute. Several MBE growth processes to obtain single crystalline $Bi_2Se_3$ films have been recently reported on a variety of substrates such as Si(111) [10], vicinal Si(111) [11], graphene [4], SiC [27], sapphire [28], GaAs [29], $SrTiO_3$ [18] etc. Among these, however, the growths on Si, the most important substrate in microelectronics, have been accompanied by disordered layers at the interface. Here, we studied the MBE growth process with different Si-substrate preparation techniques, at different substrate temperatures, and found that epitaxial $Bi_2Se_3$ films without a disordered interfacial layer can be obtained with a two-temperature-step growth. The high quality of the films was confirmed by *ex situ* high-resolution transmission electron microscopy (HRTEM), scanning tunneling microscopy (STM), X-ray diffraction (XRD), and medium energy ion scattering (MEIS).

## 2. Experimental Section

Sample Preparation and growth experiments were performed in a custom-designed SVTA MOS-V-2 MBE System [30-31]. The base pressure of the system was ~$7 \times 10^{-10}$Torr. Bi and Se fluxes were provided from Knudsen cells and the effusion cell axis was at angle of $33^o$ to the substrate normal. Bi and Se flux were measured using a quartz crystal microbalance, Inficon BDS-250, XTC/3. Si(111) substrates were un-doped with a room-temperature resistivity of $>10^3\Omega$cm. Native oxide layer was removed from the Si(111) substrates by heating using graphite filaments. The substrate temperature was calibrated at the 7x7 to 1x1 Si(111) surface reconstruction temperature of $860^o$C. The growth of the film was monitored *in-situ* by RHEED and the diffraction patterns during the growth were recorded by a charge-coupled device (CCD) camera. To ensure uniform coverage, we rotated the substrate at 5rpm during the growth and subsequent annealing in Se-atmosphere. Samples for TEM studies were prepared by focused ion-beam (FIB) technique. The TEM observations and chemical analysis were carried out with TecnaiG2F20 (200 kV) equipped with electron energy loss spectroscopy (EELS) and energy dispersive x-ray spectroscopy (EDX). STM images were taken in a custom-built low temperature STM tool using mechanically cut Pt-Ir tip at a sample bias voltage of -500mV and tunneling current of 14pA at a temperature of 12K. MEIS Data was collected using 130KeV He[+] beam and the spectrum was taken at the scattering angle of $125^o$. *Ex situ* XRD measurements were conducted using a Nonius FR571 rotating-anode generator with a copper target and graphite monochromator, giving a wavelength of 1.5418Å and detected using a Bruker HiStar multi-wire area detector.

## 3. Result and Discussion

### 3.1 Substrate Preparation

The order and cleanliness of the substrate is an important parameter to consider when establishing the optimal MBE growth conditions, since any surface contamination would result in crystal defects or second phases in the epitaxial layers [32,33]. Since the beginning of Si MBE growth, most common method employed to clean the Si substrate has been a multi-cycle high-temperature flashing process or annealing up to ~1200$^o$C [32,34-36]; and all previous reports to grow thin films on Si used such high temperature cleaning methods[10,37]. However, outgassing during the high temperature step can easily result in contamination of the reactive Si surface [32]. Thus, this method of cleaning the substrate imposes a stringent requirement of very high vacuum of the chamber. During this work, we found that clean Si(111) surface can be obtained at a much lower temperature, with an aid of an *ex situ* UV (Ultra-Violet) ozone cleaning step. Highly reactive radical oxygen formed in the presence of UV light burns off the majority of organic compounds from the surface [36]. A substrate UV cleaned for 5 minutes before mounting in the UHV chamber yielded a clean Si(111) 7x7 surface, as shown in Figure 1a, after being heated for ~2 minutes in vacuum at a temperature below 900$^o$C, which is about ~300$^o$C lower than the more common flash desorption method; considering that substrate stages rated only up to 1000$^o$C are much more common than those rated up to 1200$^o$C, this lower temperature cleaning method has a big advantage on its own.

Another MBE growth concern is the presence of dangling bonds on Si surface, which act as active sites for adsorption of contaminants [32], and the reactivity of the silicon surface itself, which leads to a high probability of formation of second phases such as SiSe$_2$ at high temperatures, leading to a rough interface [4,11,38]. To chemically saturate the dangling bonds at the Si surface and to achieve better epitaxial growth, Wu et. al. treated the interface by depositing Bi and then annealing it to form β-$\sqrt{3}$-Bi surface [10]. Besides them, many other groups have also used this technique to grow thin films on Si [11, 39,40]. Xie et. al. deposited an initial thin seed layer of $Bi_2Se_3$ at a cryogenic temperature prior to further high temperature growth, but this resulted in the formation of a few nm-thick amorphous film at the interface and twinned domains [11].

To suppress the surface effect, we implemented a very simple scheme. Inspired by the report that Se atoms can be used as a passivation layer for the surface dangling bonds on GaAs [33], we exposed the Si substrate to Se flux over a span of few seconds prior to the growth; obtaining the unreconstructed Si(111) surface; however, the substrate temperature was critical for the success of this process. At too high temperatures, $SiSe_2$ phases develop, whereas at too low temperatures amorphous selenium layer grows [41]. At a substrate temperature of ~$100^{o}$C, the deposition of Se on the Si surface was found to be self-limited without either forming thick Se or $SiSe_2$ layer; Se atoms bond only to the top Si atoms and after that no further Se gets deposited on the substrate, as verified by RHEED and TEM. Within seconds of exposure to Se, Si surface changed from 7x7 reconstruction to 1x1 structure, as observed by RHEED; and no further change in the RHEED pattern occurred even if the substrate was further exposed to Se, implying that no further reaction between Se and Si or accumulation of Se occurs at this temperature. At temperatures below ~$70^{o}$C, Se accumulation occurred, and at temperatures above ~$200^{o}$C, Se started to react with Si substrate.

3.2 Growth of $Bi_2Se_3$

The next step in optimizing our MBE growth of $Bi_2Se_3$ was to find the temperature that favors good growth and at the same time limits the decomposition of the already deposited $Bi_2Se_3$ layers. On exposing the surface to Se flux at the right temperature described above, a monolayer of Se is formed on the surface. This monolayer of Se atoms provides a passivated coverage on the Si surface and also acts as the base for the Se-Bi-Se-Bi-Se stack. Owing to the high vapor pressure of Se at relatively low temperatures [41], the first attempt was made by growing $Bi_2Se_3$ films at a comparatively low temperature. $Bi_2Se_3$ growth for ~2.5 minutes in a temperature range of $100^oC$ to $130^oC$ resulted in weak streak-like diffraction patterns, indicative of poor crystalline structure as shown in Figure 1b. Further growth at this low temperature resulted in a streaky diffraction pattern accompanied by a polycrystalline ring-like pattern, as shown in Figure 1c. Thermally annealing the sample up to ~$250^oC$ resulted in brightening the specular spot, suggesting that annealing of $Bi_2Se_3$ aids in further crystallization of the film. However, growing directly on the higher end of this temperature range resulted in an even poor quality of the film, implying that the substrate temperature was too high. In various other attempts on one-step growth, several substrate temperatures were used. As shown in Figure 1f, on depositing the initial film for ~2.5 minutes at a high substrate temperature of $350^oC$, the surface of the grown film showed 3D spot-like features along with the diffraction streaks indicating that islands of second phases such as $Bi_3Se_4$ or $SiSe_2$ coexist with the crystalline-like $Bi_2Se_3$ film. Lowering the growth temperature to $250^oC$ to deposit $Bi_2Se_3$ resulted in similar island/cluster formation, shown in Figure 1e. On lowering the temperature further to $190^oC$ and growing $Bi_2Se_3$ film for ~2.5 minutes, though no 3-D clusters were formed and sharper streaks were achieved, it was accompanied with concentric ring-like patterns shown in Figure 1d, indicating the crystallographic orientation of the film was highly disordered. Thus, we found that two-step growth temperatures were necessary to achieve second-phase-free high quality $Bi_2Se_3$ films on Si substrates right from the beginning. With a single-step high temperature growth, second phase clusters were formed, and with a low temperature growth, the crystalline quality of the films was poor even if the second phase was absent. Therefore, to avoid reaction with the silicon surface at higher temperatures, a very thin film (~2-3QL) is grown at a lower temperature ($100^oC$ - $120^oC$) at the initial stage and then to improve the crystallinity of the film, the rest of the growth is continued at a higher substrate temperature ($190^oC$ - $250^oC$). Figure 1g shows the RHEED pattern for a $Bi_2Se_3$ film grown by the two-temperature process with the high temperature growth at $190^oC$. Although the diffraction pattern shows crystalline growth, presence of faint streaks, as indicated by arrows, along with the main ones suggest that the film may have formed a twin structure, which has also been observed by other group previously [11]. However, depositing $Bi_2Se_3$ at $220^oC$ after initial low temperature growth resulted in very sharp streaky pattern, Figure 1h, indicating that the film has an atomically flat surface morphology, and no indication of twinning was observed. Further increasing the growth temperature to $300^oC$ resulted in the formation of second phase, as shown in Figure 1i. Thus, using our two-temperature growth, we were able to obtain a twin-free atomically sharp interfaces, as verified by RHEED.

Figure 2 shows the change in the spectral beam intensity during different growth steps as a 64QL film is deposited on Si at different temperatures, starting from the Se-terminated surface. It was found that at low and intermediate temperatures, the growth occurs in layer-by-layer mode, evident by the observed RHEED intensity oscillations of the diffraction beam. One period of oscillation corresponds to a deposition of one QL of $Bi_2Se_3$. It is seen in Figure 2 insets that the relative starting position for the beam intensity variation during $170^oC$ growth is the same as the end position during $110^oC$ growth, indicating that the growth indeed occurs QL-by-QL. From the RHEED oscillation period, the growth rate of $Bi_2Se_3$ QL was found to be $\sim 1QLmin^{-1}$, which was confirmed by TEM. On increasing the growth temperature to $220^oC$, RHEED oscillation is no longer observed, indicating that the growth occurs in a step-flow mode, which is a better growth mode than the layer-by-layer type. Following the low/intermediate temperature growth, the crystallinity of the film is further enhanced on annealing the sample, as indicated by brightening of the specular beam spot as sample temperature is increased to $220^oC$. The sudden jump in the intensity during annealing the film from $110^oC$ to $220^oC$ occurs around $180^oC$ showing that higher temperatures assist in crystallization of the film. Besides substrate temperature, Bi:Se flux ratio is also a key parameter in order to obtain good growth. As reported earlier by Xue et. al., it was noticed that the growth was self-limited [27,42]; i.e., growth rate was determined completely by Bi flux with excess Se species around. The Bi:Se flux ratio, measured by QCM, was kept ~1:15, with the Bi flux varied over a range from $0.87 \times 10^{13}$ cm$^{-2}$s$^{-1}$ to $2.3 \times 10^{13}$ cm$^{-2}$s$^{-1}$. The Se source was kept open at all times during the entire growth and annealing process. Within the growth temperature range, Se atoms have a non-zero coefficient only in the presence of other atoms (Bi, in this case) [41], so having the Se shutter open didn't cause any excess Se layers to be deposited on the films. Abundance of Se also minimizes the formation of Se vacancy, evident by the low carrier concentration in our films, and annealing the film in Se atmosphere further helps in crystallization of the film as observed by the brightening of the specular beam spot.

Figure 3(a-c) shows the evolution of the RHEED patterns from Se-terminated 1x1 Si surface to deposition of 1QL of Bi$_2$Se$_3$ at 110$^o$C. Except for the weak streaky pattern, any second phase or polycrystalline-ring is not observed, suggesting the onset of epitaxy right away with a single-crystal Bi$_2$Se$_3$ structure. Although the film is only 1QL thick, the spacing of the Bi$_2$Se$_3$ streaks is less than that of Si, as shown in Figure 3c. For several similar growths, the in-plane lattice constant of the 1QL film was found to be 4.14±0.05 Å, compared with 3.84 Å of Si(111) plane, which is identical to the bulk lattice constant of Bi$_2$Se$_3$. This relaxation of the Bi$_2$Se$_3$ lattice constant to its bulk value during the deposition of the first quintuple layer implies the absence of strain at the interface between the substrate and the film. The diffraction pattern becomes increasingly sharper on deposition of another QL, shown in Figure 3d. Again from RHEED, it is evident that the film quality increases as the sample temperature is increased to a higher temperature, ~220$^o$C, as can be seen in Figure 3e. The substrate temperature is ramped up slowly (~5$^o$Cmin$^{-1}$) to the higher growth temperature. It is critical to ramp the substrate temperature slowly as decomposition of the thin film has been observed on ramping the temperature too fast. Figure 3f shows the sharp twin-free RHEED pattern of a 16QL Bi$_2$Se$_3$ film grown by this two-temperature growth process. Thus, we developed a recipe with low temperature, 100$^o$C-130$^o$C, initial 2-3QL growth followed by a high temperature, 190$^o$C-220$^o$C, growth, to obtain highly crystalline, second-phase-free and atomically sharp interface between $Bi_2Se_3$ and Si substrate, as verified by RHEED, XRD and TEM.

To study the interface between the grown $Bi_2Se_3$ film and Si surface and to confirm that the growth was epitaxial right from the first quintuple layer as seen in RHEED, we conducted TEM measurements on two films grown by our two-temperature growth process. One of the samples (12 QL thick) was in atmospheric conditions for a month before TEM measurement was carried out. At the interface, an amorphous silicon dioxide layer, ~4nm thick, was observed, but this layer was formed due to diffusion of oxygen through the grown film while the sample was kept in ambient condition. EDX and EELS analysis showed that the amorphous layer is composed of only silicon and oxygen, confirming that it is not amorphous $Bi_2Se_3$, as observed by others [11]. To further verify that it is not due to the growth process, we conducted TEM measurement on another sample (32QL), which had a thick 300nm Se capping layer to prevent oxygen diffusion from air. As shown in Figure 4 (i), the cross-sectional TEM study reveals a full epitaxial structure of $Bi_2Se_3$ film down to the first quintuple layer without any amorphous layer. Both images clearly show the individual $Bi_2Se_3$ QLs with single crystalline structures. The difference in contrast of the two images is due to crystallographic orientations. Figure 4 (ii) was oriented more to emphasize the QL spacing, while Figure 4 (i) was oriented more to emphasize the Si-$Bi_2Se_3$ interface. The TEM also confirms the self-limited process of the low temperature Se deposition step on Si surface. This observation is consistent with the literature that selenide formation on Si occurs only at significantly higher temperatures [41]. By growing the initial layer at the low temperatures followed by higher temperature annealing and further growth, we were able to avoid both $SiSe_2$ and amorphous $Bi_2Se_3$ formation, getting an atomically sharp interface.

The sharp 1x1 RHEED pattern is consistent with the STM observation of atomically flat film with long terraces. From the STM image of the surface of a 100QL-thick $Bi_2Se_3$, Figure 5, it is seen that triangular terraces are formed, reflecting the three-fold symmetry of the bulk $Bi_2Se_3$ along hexagonal [001] direction and each terrace step is ~1nm, i.e. 1QL along the c-axis, indicating that the growth occurs 1QL-by-1QL. The chemical stoichiometry was confirmed by MEIS for the grown $Bi_2Se_3$ films as thin as 3QL. Figure 6 shows MEIS data for a thin 5QL $Bi_2Se_3$ film; the films were found to be stoichiometric with Bi:Se = 2.0:3.0. XRD (X-Ray Diffraction) provides the quintuple layer spacing of ~9.6Å, consistent with that of bulk samples, as shown in Figure 7. By inspecting the details of the XRD peaks, we see that our thin films, as thin as 4QL, are c-axis oriented along the growth direction and have the same structure as bulk samples.

## 4. Conclusions

To summarize, we found that the key parameters for obtaining single crystalline $Bi_2Se_3$ film on silicon substrates without any interfacial layers are the substrate preparation prior to growth and the two-temperature growth process. Contrary to the common belief found in literature that either Bi, amorphous $Bi_2Se_3$ or some other buffer layer is required to grow single-crystalline $Bi_2Se_3$ heteroepitaxially on Si, we have shown that these films can be grown with an atomically abrupt interface on Si without any interfacial layers using our two-temperature growth process. The high quality of MBE-grown $Bi_2Se_3$ films with precise layer thickness control and interfacial abruptness on Si(111) substrate will make it a strong candidate for further study of topological insulators. Incorporation of the topological insulators into the well-established Si-industry will open new horizons for spintronics and quantum computational devices.

## Acknowledgements
This work is supported by IAMDN of Rutgers University, National Science Foundation (Grant No. NSF DMR-0845464) and Office of Naval Research (Grant No. ONR N000140910749).

**Figure Captions**

**Figure 1.** RHEED patterns for different growth procedures. a) Si 7x7 surface obtained by UV-ozone cleaning and heating in UHV chamber. b) Low temperature growth at $110^{o}$C for 2.5 minutes. c) Continued low temperature growth for another 10 minutes. d-f) Initial growth of ~2.5 minutes at high temperatures of $190^{o}$C, $250^{o}$C and $350^{o}$C, respectively. g-i) Two temperature growth with initial 2-3QL deposition at $110^{o}$C and higher temperature growth at $190^{o}$C, $230^{o}$C and $300^{o}$C, respectively.

**Figure 2.** Evolution of the specular beam intensity during $Bi_2Se_3$ growth at different temperature. After 2QL deposition is done at $110^{o}$C, the film is slowly heated to $220^{o}$C. The temperature is reduced to $170^{o}$C and another 4QL are deposited. In the inset, clear RHEED oscillations can be seen for $170^{o}$C growth and even for $110^{o}$C growth. Specular beam intensity again increases on annealing to $220^{o}$C, where it slowly increases during the 58QL deposition without RHEED oscillation, implying that the growth occurs through step-flow mode at this temperature. Further increase in the specular beam intensity is observed on cooling the sample in Se atmosphere, indicating the Debye-Waller effect.

**Figure 3.** Evolution of RHEED pattern on the surface from Si(111) 1x1 structure obtained by Se exposure at low temperature to the $Bi_2Se_3$ layers. a) Surface after a monolayer of Se makes bonding with Si and removes the 7x7 reconstruction. b) Structure midway the first QL deposition after ~0.5QL of $Bi_2Se_3$. c) After deposition of 1QL of $Bi_2Se_3$. Streaky pattern indicative of single-crystalline structure can be seen. d) The diffraction pattern improves on deposition of another QL. e) RHEED pattern gets much brighter and sharper after annealing the 2QL $Bi_2Se_3$ film to $220^{o}$C. f) Final RHEED pattern of a 16QL film grown by the two-temperature growth process.

**Figure 4.** (i) TEM cross-section showing the sharp interface between the $Bi_2Se_3$ film and the Si(111) substrate for the sample with a 300nm thick Se-capping. (ii) TEM cross-section of a $Bi_2Se_3$ sample without a capping layer, which was kept in ambient condition for weeks before the TEM measurement was done. Due to oxygen diffusion through the grown film, an amorphous layer of silicon dioxide is formed at the interface, as verified by EDX and EELS. Due to the difference in sample orientations, contrast is clearer for silicon ordering and interface in (i) and $Bi_2Se_3$ QLs in (ii). The spacing between two consecutive dotted lines represents 1QL, which is about 1 nm thick. The samples are prepared with the standard Ga-ion FIB (Focused Ion Beam) process.

**Figure 5.** STM image of a 100QL sample grown on Si(111) substrate. The triangular terraces are indicative of the tri-fold symmetry of the $Bi_2Se_3$ (111) plane.

**Figure 6.** MEIS data for a 5QL thick $Bi_2Se_3$ film. From MEIS, we see that our films are stoichiometric with the composition of Bi:Se = 2.0:3.0

**Figure 7.** XRD measurements for $Bi_2Se_3$ thin films of different thicknesses; the peak position provides quintuple layer spacing of ~9.6Å along the c-axis.

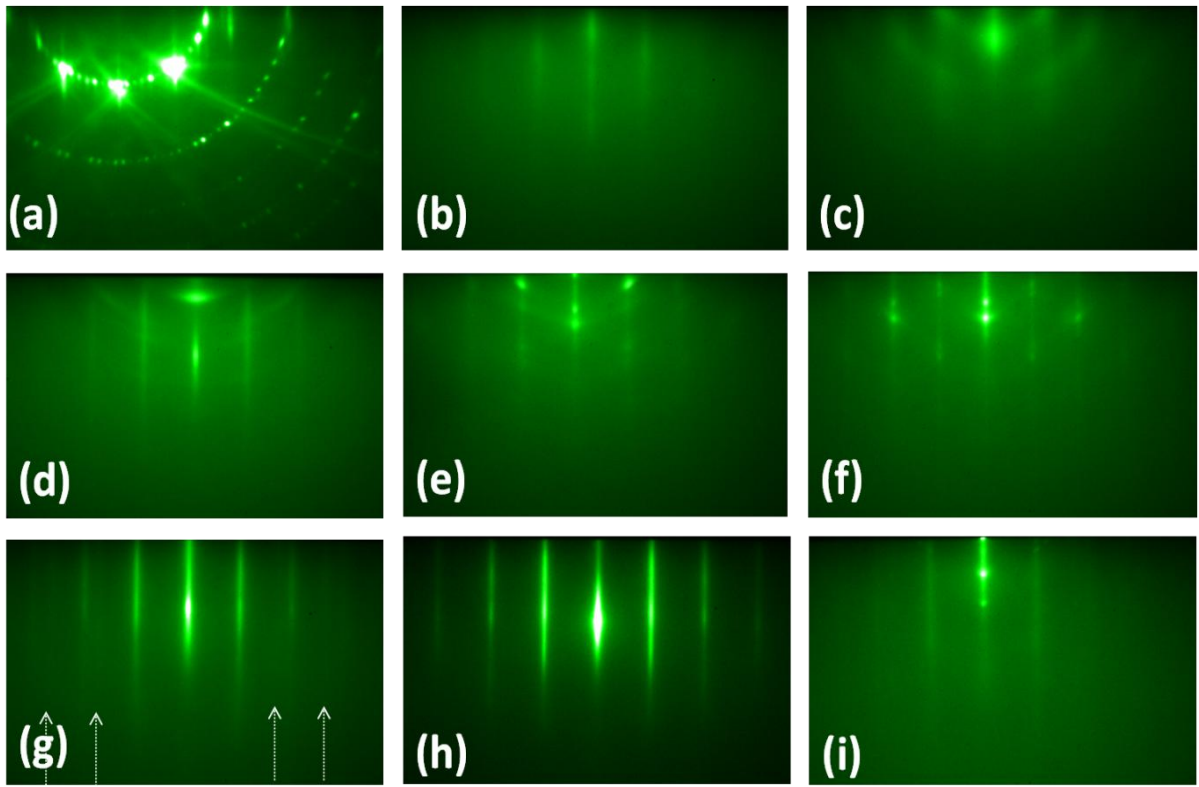

Figure 1.

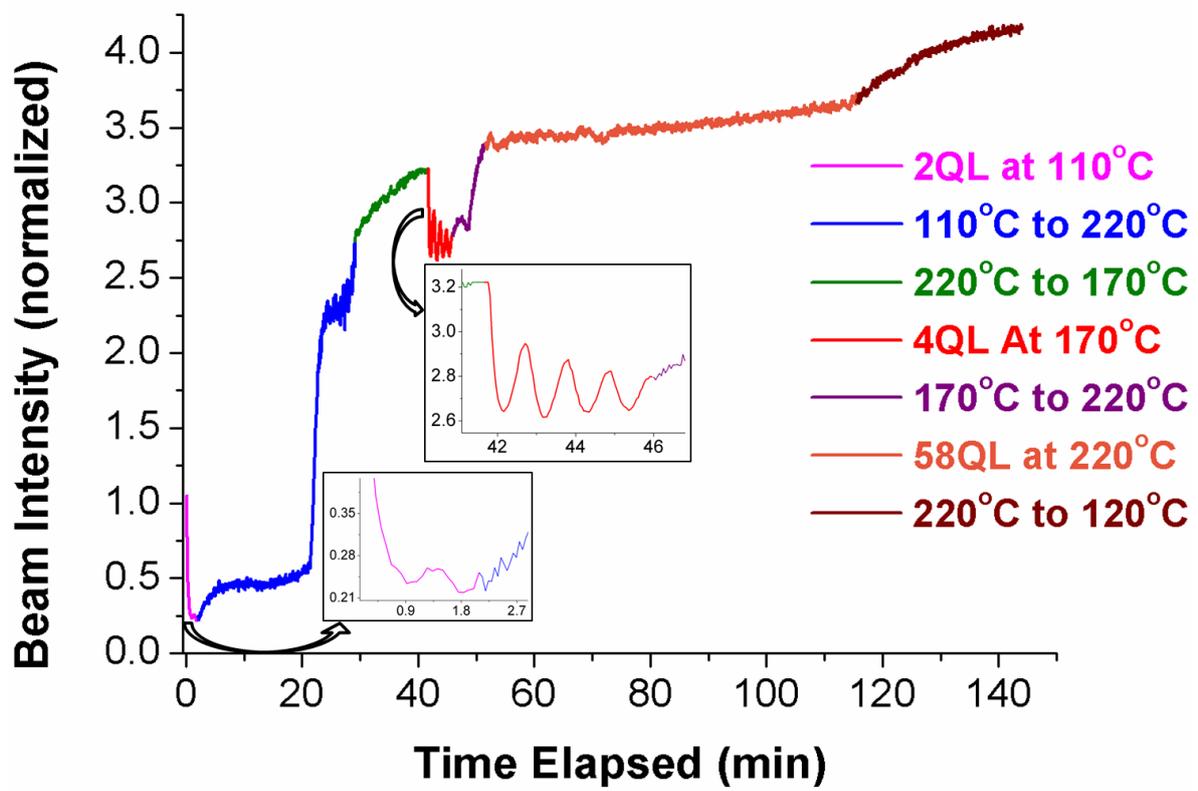

Figure 2.

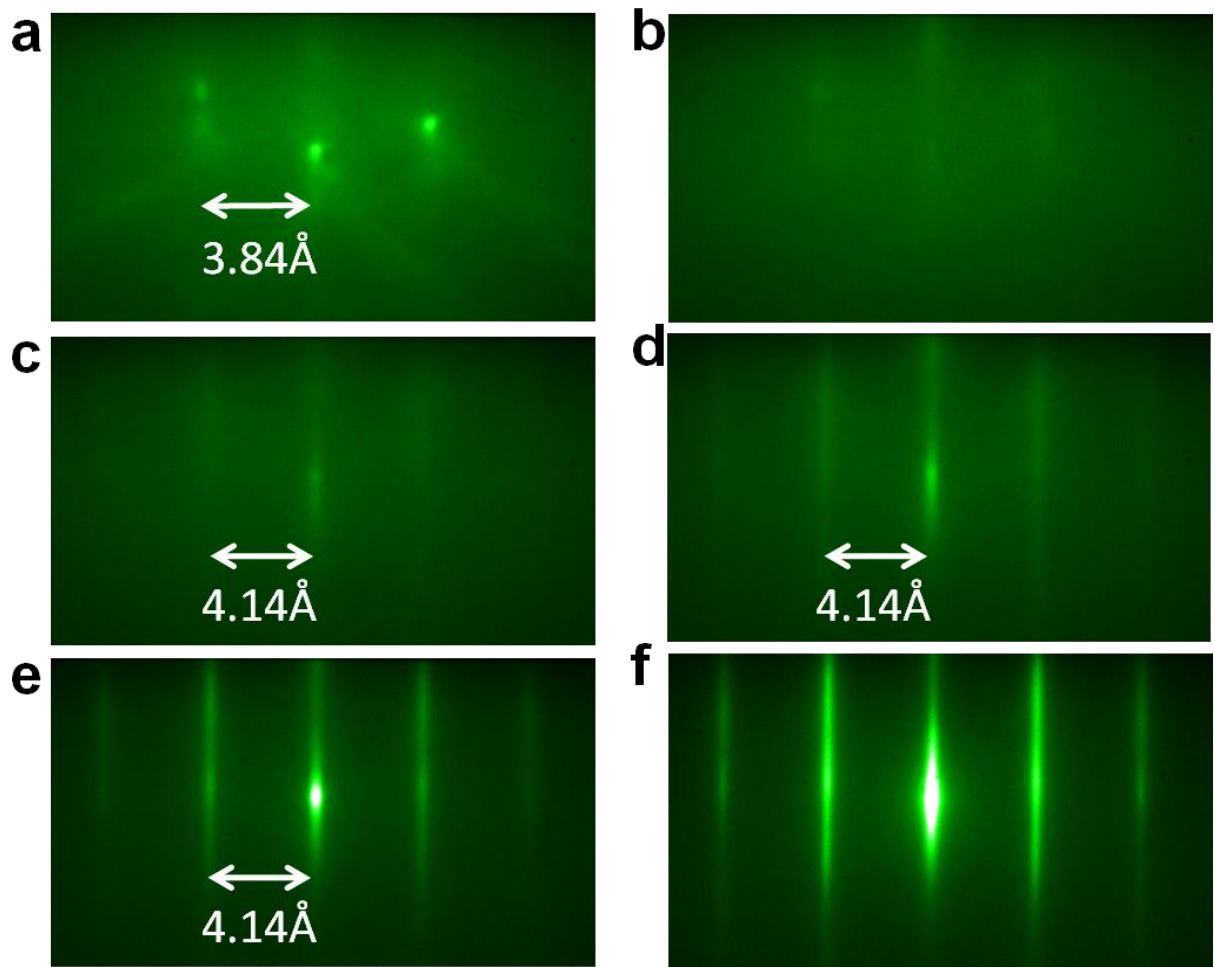

Figure 3.

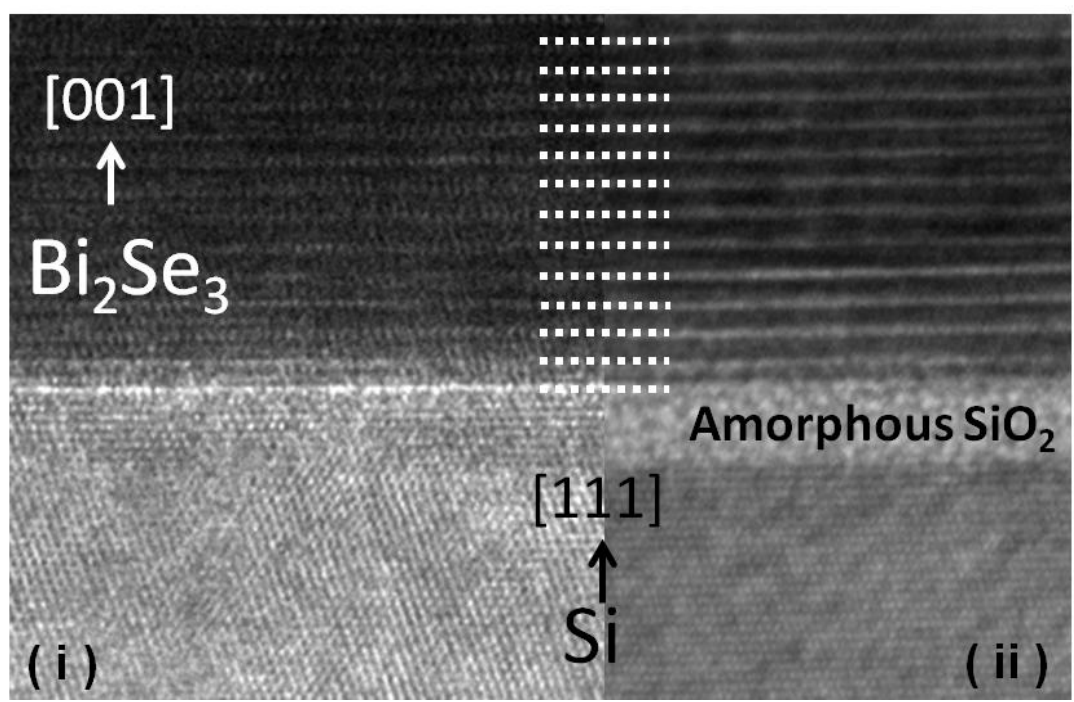

Figure 4.

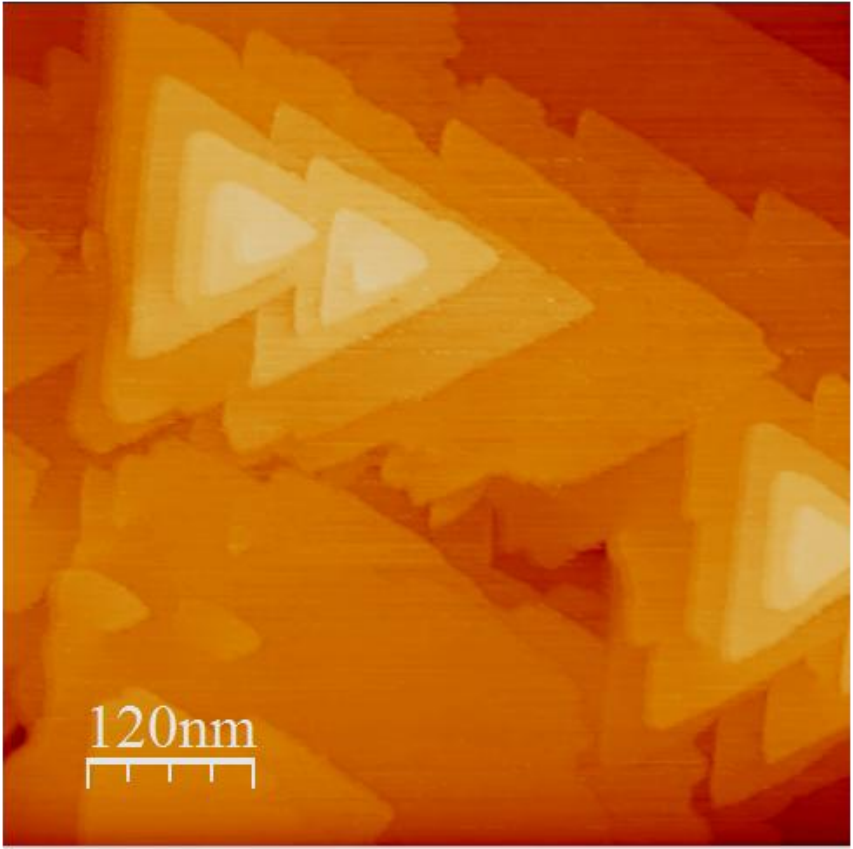

Figure 5.

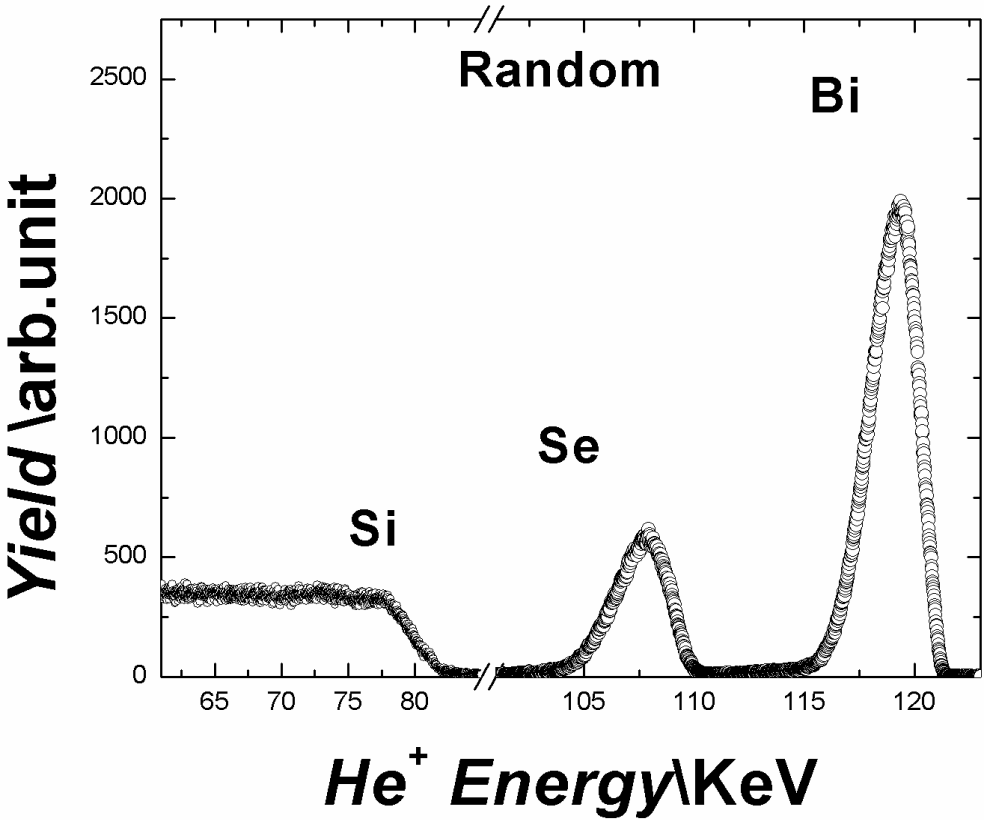

Figure 6.

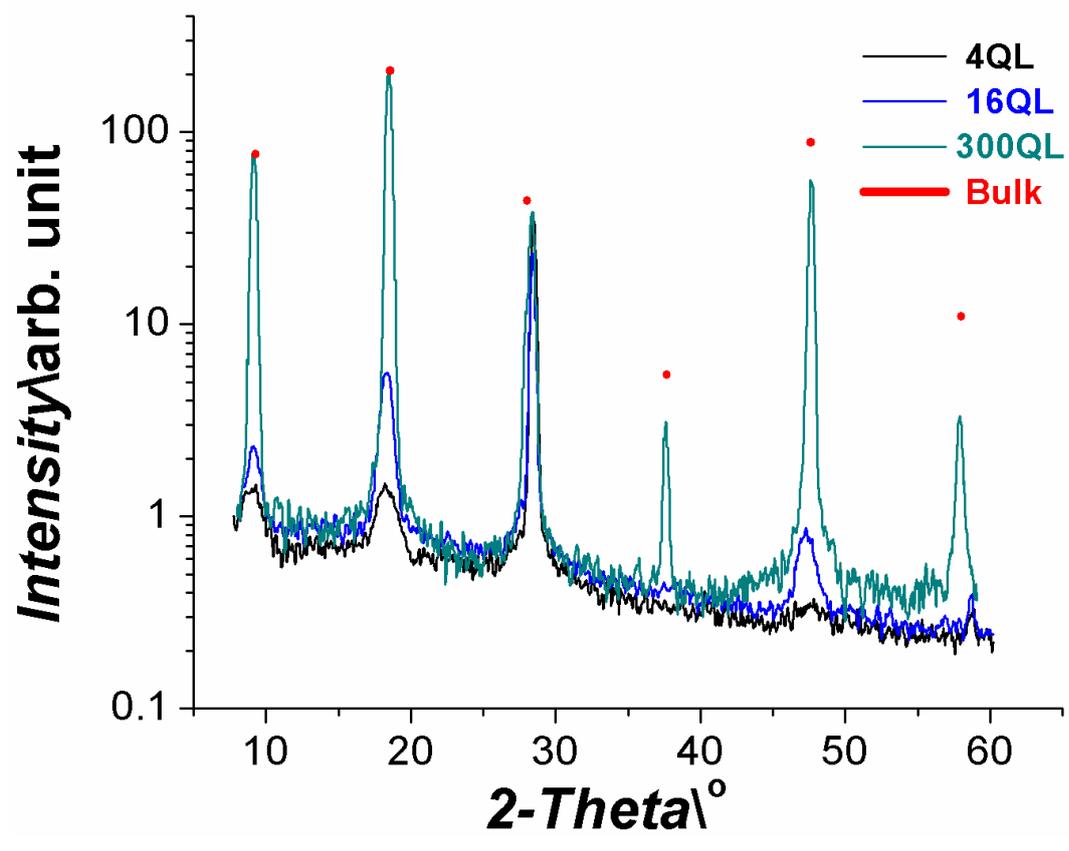

Figure 7.